\documentclass[prd,aps,twocolumn,floats,floatfix,nofootinbib] {revtex4-1}
\pdfoutput=1

\usepackage[T1]{fontenc} % if needed

\usepackage{graphicx}
\usepackage{dcolumn}
\usepackage{bm}
\usepackage{graphics}
\usepackage{slashed}
\usepackage{amssymb}
\usepackage{natbib}
\usepackage{amsmath}
\usepackage{url}
\usepackage[usenames,dvipsnames,svgnames,table]{xcolor}
\usepackage{color}
\usepackage{xcolor}
\usepackage{tabularx}
\usepackage{hyperref}
\usepackage[utf8]{inputenc}

\newcommand{\K}{\mathcal{K}}

\newcommand{\OO}{\mathcal{O}}

\newcommand{\CC}{\mathcal{C}}

\newcommand{\E}{\mathcal{E}}

\newcommand{\no}{\noindent}
\newcommand{\nb}{\nonumber}
\def\L{\Lambda}
\def\l{\lambda}
\def\m{\mu}
\def\n{\nu}
\def\p{\phi}
\def\na{\nabla}

\def\b{\beta}
\def\g{\gamma}
\def\s{\sigma}

\def\de{\partial}
\def\beq{\begin{equation}}
\def\eeq{\end{equation}}
\newcommand{\bea}{\begin{eqnarray}}
\newcommand{\eea}{\end{eqnarray}}
\newcommand{\be}{\begin{equation}}
\newcommand{\ee}{\end{equation}}

\begin{document}
\title{\boldmath K-dynamics: well-posed 1+1 evolutions in K-essence}
\author{Miguel Bezares$^{1,2}$, Marco Crisostomi$^{1,2}$, Carlos Palenzuela$^{3}$ and Enrico Barausse$^{1,2}$\\$\,$}

\affiliation{$^{1}$SISSA, Via Bonomea 265, 34136 Trieste, Italy and INFN Sezione di Trieste \\
$^{2}$IFPU - Institute for Fundamental Physics of the Universe, Via Beirut 2, 34014 Trieste, Italy \\
$^{3}$Departament de F\'isica \& IAC3, Universitat de les Illes Balears, Palma de Mallorca, Baleares E-07122, Spain}

\begin{abstract}
We study the vacuum Cauchy problem for K-essence, i.e. cosmologically relevant scalar-tensor theories  that involve 
first-order derivative self-interactions,
and which pass all existing gravitational wave bounds. 
We restrict to spherical symmetry and show that there exists a large class of theories for which no breakdown of the Cauchy problem
occurs outside apparent black hole horizons, even in the presence of scalar shocks/caustics, except for a small set of initial data sufficiently close to critical black hole collapse.
We characterise these problematic initial data, and show that they lead to large or even diverging (coordinate) speeds for the characteristic curves.
We discuss the physical relevance of this problem and propose ways to overcome it.
\end{abstract}

\date{\today}

\maketitle
\flushbottom

\section{Introduction}
Supplementing General Relativity (GR) with an additional scalar  degree of freedom 
yields a simple and popular class of gravitational theories: scalar-tensor theories. These
extensions  of GR can be divided into two broad sub-classes: theories
without derivative self-interactions (e.g. Fierz-Jordan-Brans-Dicke theories \cite{Fierz:1956zz,Jordan:1959eg,Brans:1961sx}); and ones with derivative self-interactions of the scalar field.
The latter, when restricted to 
actions containing at most two derivatives for each occurrence of the scalar field, have been recently classified in the framework of degenerate higher-order scalar-tensor (DHOST) theories \cite{Langlois:2015cwa, Crisostomi:2016czh, BenAchour:2016fzp} (which include Horndeski \cite{Horndeski:1974wa} and beyond Horndeski \cite{Gleyzes:2014dya} theories). When the theory is in the perturbative regime, the derivative self-couplings are not expected to be important, since they are suppressed by powers of the ultra-violate (UV) cutoff scale. However, in the non-linear regime those operators 
produce  a novel and important effect, i.e. the {\it screening} of local scales
from GR modifications (see \cite{Joyce:2014kja} for a review). 
This screening is crucial to pass local experimental tests (at the level of the solar system)~\cite{Will:1993hxu,Will:2014kxa},
while allowing for significant effects on cosmological scales.
Moreover, several subsets of DHOST theories have been proven to be radiatively stable \cite{Santoni:2018rrx}, so that they can develop large non-linearities while remaining quantum-mechanically stable. 

In this paper, we focus on theories with action depending only
on first derivatives of the scalar field.
In the absence of gravity, these are known as $P(X)$ theories, where
 $X\equiv (\de \phi)^2$ is the kinetic term of the scalar field
and $P$ its Lagrangian density. However, 
when gravity is taken into account,  they are usually called {\it K-essence} theories.  A first nice property of such actions
is that they provide second order equations of motion, and therefore they are automatically free from Ostrogradsky instabilities \cite{Ostrogradsky:1850fid}.
Secondly, they enjoy a kinetic screening mechanism on local scales, known as {\it K-mouflage} \cite{Babichev:2009ee}. K-essence theories have been widely applied both in early and late time cosmology. They were introduced in the context of K-inflation \cite{ArmendarizPicon:1999rj}, and then used to explain the present accelerated expansion of the Universe (self-acceleration) \cite{Chiba:1999ka, ArmendarizPicon:2000dh}. 
Particular functions of $X$ attracted special attention, e.g. the Dirac--Born--Infeld (DBI) models \cite{BornInfeld, Dirac}, which arise in the action for D-branes in string theory, and the ghost condensate \cite{ArkaniHamed:2003uy}.
Finally, it is worth noting that the recent  GW170817 bound on the speed of 
tensor modes~\cite{Monitor:2017mdv,TheLIGOScientific:2017qsa}, combined 
with other constraints also based on the propagation of gravitational waves \cite{Creminelli:2018xsv, Creminelli:2019kjy, Babichev:2020tct}, 
leaves K-essence (with a conformal coupling to matter) unconstrained. K-essence is therefore one of the most compelling alternatives to Dark Energy.

However, it is still unclear whether the initial data (i.e. Cauchy) problem is well-posed in K-essence, i.e., 
according to Hadamard's definition~\cite{hadamard}, whether it admits a unique solution that depends continuously on the initial data.
Understanding this point is of paramount importance  if we want to seriously consider K-essence to describe real astrophysical objects such as (neutron) stars or black holes, in isolation or in binaries. Recently, several papers have tackled this problem by using numerical simulations \cite{2011JHEP...04..096A,2011JHEP...10..028L,Bernard:2019fjb}, concluding that well-posedness is not always guaranteed. However, a quantitative estimation of the regime of well-posedness is still unavailable (see \cite{Figueras:2020dzx} for a recent attempt). Moreover,
 caustics in the scalar field characteristics, and therefore scalar shocks, are known to form generically
in these theories~\cite{Felder:2002sv, Reall:2014sla,Babichev:2016hys,Babichev:2017lrx,Bernard:2019fjb}, which makes numerical evolutions even more challenging. 

In this paper, we show that a broad class of K-essence theories yields a
strongly hyperbolic and thus  well-posed Cauchy problem in vacuum and spherical symmetry, for initial
data sufficiently far from critical black hole collapse~\cite{Choptuik:1992jv,Gundlach:1999cu}. With the exception
of this (small) set of initial data (which we identify with a precise criterion), we manage to avoid 
most of the pathologies previously found in the literature \cite{2011JHEP...04..096A,2011JHEP...10..028L,Bernard:2019fjb,Figueras:2020dzx},  at least
outside apparent black hole horizons.
Moreover, we show that one can write the scalar equation as a conservation law, which allows
for using shock capturing methods  to evolve the scalar field.
As for the problematic initial data close to critical black hole collapse, we find that the evolution equations become of the 
Keldysh type~\cite{Ripley2019,Bernard:2019fjb} during the evolution, i.e. the characteristic velocities become very large or even diverge.
While in principle this problem may be fixed  with a different gauge choice (e.g. by introducing a non-zero shift), we
could not find a suitable scheme fixing it.

The paper is organised as follow: in Sec.~\ref{PofX} we review the key features of $P(X)$ theories, i.e. K-essence in the absence of gravity; this is sufficient to recap the most important properties and issues of this class of theories without the complication of gravity. In Sec.~\ref{Kessence} we reintroduce gravity, set the stage to perform numerical evolutions, and discuss our criterion to establish the regime of well-posedness. In Sec.~\ref{NumSetup} we present our numerical setup, and in Sec.~\ref{results} we show the results of our simulations. Finally, in Sec.~\ref{Conc}, we draw our conclusions. We use the 
$(-+++)$ signature, denote spacetime (space) indices with Greek (Latin) letters, and set $c=1$.

\section{$P(X)$ theories: a short review}
\label{PofX}

$P(X)$ actions involve a single derivative for each occurrence of the scalar field, and are invariant under the shift symmetry $\phi(x) \to \phi(x) + c$, where $c$ is  a constant.
The lowest-order Lagrangian is therefore given by
\be
L= - \frac12 X + \frac{\b}{4 \L^4} X^2 + a\sqrt{G} \phi \, T \,, \label{actionPofX}
\ee
where $X\equiv (\de \phi)^2$,  $\b$ and $a$ are dimensionless coefficients of $\sim \OO(1)$, 
$G$ is the gravitational constant that sets the strength of the coupling with the trace of the matter energy-momentum tensor $T$,
and $\L$ is the strong coupling scale of the effective field theory (EFT)\footnote{Note that in our units $c=1$, $\phi$ has dimensions $[G^{-1/2}]$, $\L^4$ has dimensions $[E/L^3]$, with $E$ an energy and $L$ a length.}.
The equation of motion  reads
\be
\Box \phi - \frac{\b}{\L^4} \de_\m \left[ (\de \phi)^2 \de^\m \phi \right] + {a\sqrt{G}} T = 0 \,.
\ee

\subsection{Kinetic screening}
\label{screening}

Let us now assume a static and spherically symmetric configuration $\phi = \phi (r)$, and a point-like matter source $T = - m \, \delta (r)$ of mass $m$. Thanks to the shift symmetry, the equation of motion can be integrated once and gives
\be
\phi' - \frac{\b}{\L^4} {\phi'}^3 = \frac{a \sqrt{G}\, m}{4 \pi r^2} \,. \label{screeneq}
\ee
This equation can be solved analytically for $\phi'$, 
but since it is a cubic polynomial the solution is not particularly illuminating.
It is in fact  more useful to look at it in two extreme regimes.

\begin{itemize}
\item Far away from the source:

In this regime $\phi'/\L^2 \ll 1$, and therefore the cubic term in Eq. (\ref{screeneq}) is negligible compared to the linear one.
The solution reads
\be
\phi'_{\text{lin}} \simeq \frac{a \sqrt{G}\, m}{4 \pi r^2} \,, \label{philin}
\ee
and we recover the standard quadratic fall-off of the ``scalar force'' $\phi'$.

\item Close to the source:

In this regime $\phi'/\L^2 \gg 1$, and it is the cubic term in Eq.~(\ref{screeneq}) that dominates over the linear one. Note that a solution in this case is only possible if $\beta < 0$, in which case one obtains
\be
\phi'_{\text{non-lin}} \simeq \left( - \frac{\L^4}{\b} \frac{a \sqrt{G} \, m}{4 \pi r^2} \right)^{1/3} \,. \label{phinonlin}
\ee
\end{itemize}

\no The radius where the transition between the linear and non-linear regimes occurs can be easily obtained from the equation of motion (\ref{screeneq}), as the radius where the cubic term becomes comparable to the linear one, i.e. $\phi' \simeq \L^2$. By inserting this value into Eq.~(\ref{screeneq}), one obtains, up to factors of order one,
\be
r_k \simeq \frac{1}{\L} \left( \sqrt{G} m \right)^{1/2} \,.
\ee
Expressing the solutions (\ref{philin}) and (\ref{phinonlin}) in terms of $r_k$, we find the ratio
\be
\frac{\phi'_{\text{non-lin}}}{\phi'_{\text{lin}}} \propto \left( \frac{r}{r_k} \right)^{4/3} \,.
\ee
Therefore, when $r \ll r_k$ the scalar force is highly suppressed with respect to what it would have been without the $(\de \phi)^4$ term in the Lagrangian. This is the essence of the kinetic screening of the scalar force.
This feature is not a mathematical curiosity, but has strong physical implications. It allows for a sensible modification of gravity at large (cosmological) distances, and at the same time suppresses the modification at local scales, where GR is very well tested. This is impossible with Fierz-Jordan-Brans-Dicke like theories, where the deviations away from GR  appear at all scales.

Moreover, this simple calculation shows how the non-linear regime of $P(X)$ is much more interesting than the linear one. In the latter, the various powers of $X$ in the Lagrangian are simply ``irrelevant''.

\subsection{Issues and solutions}
\label{issues}

It is fair at this point to mention that the nice properties  of the kinetic screening 
that we reviewed above are just one side of the story. Indeed, kinetic screening also presents
some problematic aspects. 

First of all, the screening starts to be effective when $X/\L^4 \gtrsim 1$. This is exactly where the EFT expansion seems to break down. In this regime, one would expect higher and higher powers of $X$ to become more and more important, and it would not
be justified to neglect them in the Lagrangian. Also, higher derivatives terms (with more than one derivative for each occurrence
of the scalar field in the action) are generically expected to be produced by quantum corrections. Note however that Ref.~\cite{deRham:2014wfa} 
showed that when radiative corrections are correctly computed, any given form of $P(X)$ is radiatively stable in the non-linear regime. Ref.~\cite{Brax:2016jjt} confirmed this result also in presence of gravity, following a different approach.

Second, the choice $\b < 0$, necessary for screening, makes the perturbation of the scalar field propagate at superluminal speed. This can also be seen as the failure to satisfy the analyticity requirement of the $2 \to 2$ scattering amplitude in the forward limit, which requires $\b > 0$ \cite{Adams:2006sv}. This suggests that $P(X)$ theories with screening cannot arise as the low energy EFT of a Lorentz invariant and local theory. However,  it has been shown very recently that the presence of gravity can substantially change this bound on $\beta$ \cite{Alberte:2020jsk}.

Finally, it was noted in \cite{Babichev:2016hys} that $P(X)$ theories lead to caustic formation when scalar waves propagate. However, it should be stressed that caustics/shocks are not necessarily a problem, as they happen also in realistic systems where the weak or integral solution is considered instead of the strong or classical one (e.g. in hydrodynamics). In order to deal with them numerically, we cast the scalar equation into  a conservation law form, and use high-resolution shock-capturing methods to solve them.

\section{K-essence and well-posedness}
\label{Kessence}

Let us now include gravity, and consider the dynamics of K-essence in vacuum. 
The action is given by
\be
S=\int d^{4}x\sqrt{-g}\left[\frac{R}{16 \pi G} + K(X)  \right]  \,, \label{action}
\ee
where $R$ is the Ricci scalar of the spacetime metric $g_{\mu\nu}$, $K(X)$ is a function of the kinetic term of the scalar field~$\phi$, 
and like before we define $X \equiv \nabla_{\mu}\phi \nabla^{\mu} \phi$.
Here, we will consider only the lowest order terms
\be
K(X) = - \frac12 X + \frac{\beta}{4\, \Lambda^4}X^{2} - \frac{\gamma}{8\,\Lambda^8}X^{3} \,, \label{Kform}
\ee
where $\L$ is the strong coupling scale of the EFT and $\b, \, \g$  are dimensionless coefficients of $\sim \OO(1)$.

The equations of motion for the metric and the scalar field are respectively 
\begin{eqnarray}
 &&G_{\mu\nu} = 8 \pi G T^{\phi}_{\mu\nu} \,,\label{EE}\\[2ex]
 &&\nabla_{\mu} (K'(X) \, \nabla^{\mu}\phi) = 0 \,, \label{KG}
\end{eqnarray}
where $T^{\phi}_{\mu\nu} = K(X)  \, g_{\mu\nu} -2\,  K'(X) \, \na_\m \p \na_\n \p,$ and primes denote derivatives with respect to $X$. 
The scalar field equation can also be recast into a generalised Klein-Gordon equation  $\gamma^{\mu\nu}\nabla_{\mu}\nabla_{\nu}\phi= 0$, with an effective metric
\be
\gamma^{\mu\nu} \equiv g^{\mu\nu} + \frac{2\,K''(X)}{K'(X)}\nabla^{\mu}\phi\nabla^{\nu}\phi \,. \label{effmetric}
\ee

\subsection{Evolution equations in spherical symmetry}

In order to investigate the non-linear dynamics of $K(X)$ theories, it is natural to first restrict the analysis to the spherically symmetric case. 
To write the equations of motion as an evolution  system, the spacetime tensors and equations can be split into their space and time components by adopting a $1+1$ decomposition. The line element can be 
decomposed as
\begin{equation}\label{metric_ansatz}
ds^2 = -\alpha^2(t,r)dt^2+g_{rr}(t,r)dr^2+r^2g_{\theta\theta}(t,r)d\Omega^2~,
\end{equation}
where $\alpha$ is the lapse function, $g_{rr}$ and $g_{\theta\theta}$ are positive metric functions, and 
$d\Omega^2=d\theta^2+\sin^2\theta d\varphi^2$ is the solid angle element. These quantities are defined
on each spatial foliation $\Sigma_{t}$ with normal $n_{a}=(-\alpha,0)$ and extrinsic curvature $\K_{ij} \equiv 
-\frac{1}{2}\mathcal{L}_{n}\gamma_{ij}$, where $\mathcal{L}_{n}$ is the Lie derivative along $n^{a}$. 
Note that this rather generic choice of coordinates allows us to follow the dynamical evolution not only up to black hole formation, but also
past it.

The Einstein equations \eqref{EE} can be written as an evolution system by using the Z3 formulation in spherical symmetry~\cite{Alic:2007ev,Bona:2005pp,Bernal:2009zy,ValdezAlvarado:2012xc}. We can express \eqref{EE}-\eqref{KG} as a 
first order  system by introducing first derivatives of the fields as independent variables, namely
\begin{eqnarray}
A_{r}=\frac{1}{\alpha}\partial_{r}\alpha \,,  \quad  &&{D_{rr}}^{r}=\frac{g^{rr}}{2}\partial_{r}g_{rr} \,, \quad  {D_{r\theta}}^{\theta}=\frac{g^{\theta\theta}}{2}\partial_{r}g_{\theta\theta} \,,\nonumber \\
\Phi&=&\partial_{r}\phi \,, \qquad \Pi=-\frac{1}{\alpha}\partial_{t}\phi \,.
\end{eqnarray}
The resulting evolution system can be written as a system of conservation equations
\begin{equation}
\partial_{t}{\bf U} + \partial_{r} F( {\bf U} ) = \mathcal{S}( { \bf U}) \,, 
\end{equation}
where ${\bf U} 
=\{\alpha\,, g_{rr}\,, g_{\theta\theta}\,, {\K_{r}}^{r}\,, {\K_{\theta}}^{\theta}\,,A_{r}\,, {D_{rr}}^{r}\,, {D_{r\theta}}^{\theta}\,, Z_{r}\,, \\ \phi\,, \Phi\,, \Pi \}$ is a vector containing the full set of evolution fields, $Z_{r}$ is the time integral of the momentum constraint\footnote{In the Z3 formulation, the momentum constraint is included into the evolution system by considering an additional vector $Z_{i}$ as an evolution field~\cite{ValdezAlvarado:2012xc}.}, $F( {\bf U} )$ is the radial flux and $\mathcal{S}( { \bf U})$ is a source term. The  evolution equations for the Z3 formulation can be found explicitly in  Ref.~\cite{ValdezAlvarado:2012xc}. A gauge condition for the lapse is required to close the system. We use the singularity-avoidance $1+\log$ slicing condition $\partial_{t}\ln\alpha=-2\,\mathrm{tr}\K,$ 
where $\mathrm{tr}\K=\K_{r}^{r}+2\K^{\theta}\,_{\theta}$~\cite{BM}.
Other singularity-avoidance prescriptions have been used, leading to similar results.

The equation of motion~\eqref{KG} for the scalar field, written in conservative form, becomes
\begin{eqnarray}
\partial_{t}\phi &+& \alpha\Pi = 0\,,\label{KG2-1}\\
\partial_{t}\Phi &+& \partial_{r}\left[\alpha\Pi \right] = 0\,,\label{KG2-2}\\
\partial_{t}\Psi &+& \partial_{r}F_{\Psi}=-\frac{2}{r}F_{\Psi} \,,\label{KG2-3}
\end{eqnarray}
where
\bea
\Psi&=&\sqrt{g_{rr}}g_{\theta\theta} K' \Pi\,, \label{fPi} \\[2ex]
F_{\Psi}&=&\alpha\sqrt{g_{rr}}g_{\theta\theta}K' g^{rr}\Phi \,.
\eea
Note that we have introduced a new ``conservative'' field~$\Psi$, which depends implicitly on the primitive fields $\{\Pi,\Phi\}$ through the non-linear equation~\eqref{fPi}. During the evolution, this equation
needs to be solved numerically at each time-step to recover $\Pi$, as the  latter appears in the evolution system  (\ref{KG2-1}-\ref{KG2-2}) and in the stress-energy tensor in the Einstein equations. 

\subsection{Character and velocities}

In order to assess the well-posedness of the Cauchy problem in K-essence, we first analyse the character
of the evolution system. More specifically, we aim to ascertain whether the system is strongly
hyperbolic,  since  that is  a sufficient  condition for a well-posed Cauchy problem~\cite{10.1007/3-540-46580-4_2,Reula:1998ty}.

There are many different approaches to determine the hyperbolicity of a system, most of which 
result in  algebraic conditions that  the system  needs to satisfy~\cite{Sarbach:2012pr}. When the equations are quasilinear (like in our case), 
the system is strongly hyperbolic if its principal part (i.e. 
the system obtained by considering only the highest derivatives)
 has real eigenvalues and a complete set of eigenvectors\footnote{For weakly hyperbolic systems (i.e.~when the principal part has real eigenvalues, but an incomplete set of eigenvectors) the Cauchy problem is not necessarily well-posed, and the solution may grow exponentially, which would  give an unstable numerical evolution.} \cite{Hilditch:2013sba}.

For our scalar field evolution  system (\ref{KG2-1}-\ref{KG2-3}), the characteristic matrix
for the principal part is
\begin{equation}
\mathbb{M}= 
\left(\begin{array}{cc} 
0 & \frac{\alpha}{\sqrt{g_{rr}}}\\[2ex]
-\frac{\sqrt{g_{rr}}}{\alpha}\frac{\gamma^{rr}}{\gamma^{tt}} \,\,\,\,& -\frac{2\, \gamma^{tr}}{\gamma^{tt}} 
\end{array}\right) \,.
\end{equation}
The eigenvalues of this  matrix, $V_{\pm}$, read
\begin{eqnarray}
V_\pm = -\frac{\gamma^{tr}}{\gamma^{tt}}\pm \sqrt{\frac{-{\rm det}(\gamma^{\mu\nu})}{(\gamma^{tt})^2}} \,, \label{Vpm}
\end{eqnarray} 
where  $\gamma^{\mu\nu}$
is given by Eq.~(\ref{effmetric}), or more explicitly by
\begin{eqnarray}\label{gamma1}
\gamma^{tr}&=& \frac{2\,K''\,\Pi\,\Phi}{\alpha\, g_{rr}\, K'} \,, \\
\gamma^{tt} &=& - \frac{1}{\alpha ^2} \left( 1 - \frac{2\, K''}{K'}\Pi^2 \right)\,, \\
\gamma^{rr} &=& \frac{1}{g_{rr}} \left( 1 + \frac{2\, K''}{K'} \frac{\Phi ^2}{g_{rr}} \right) \,.\label{gamma3}
\end{eqnarray}
The criterion given above for strong hyperbolicity is then satisfied if $V_\pm$ are real and distinct, i.e. 
if ${\rm det}(\gamma^{\mu\nu})<0$.

Note that $V_{\pm}$, which can be physically interpreted
as \textit{coordinate} ``characteristic speeds'' (i.e. the radial propagation speeds of the scalar  wavefronts
in the geometric optics limit),
are functions of both space and time through $\gamma^{\mu\nu}$.
This can potentially give rise to shocks even from  smooth initial data, which may lead to
non-unique solutions and thus to an ill-posed Cauchy problem~\cite{Reall:2014sla,evans10,Tanahashi:2017kgn}.
Our set-up avoids this issue by writing the scalar evolution equations
as a hyperbolic conservation system, c.f. Eqs.
(\ref{KG2-1}-\ref{KG2-3}),
which we solve using high-resolution shock capturing
techniques (as far as the weak solution, corresponding to the integral version of the equations, is unique).

Note also that $V_{\pm}$ provides the non-linear generalization to the usual (linear) expression for the speed
of sound in K-essence, $c_s=\pm\sqrt{1+2XK''/K'}$ (see e.g. \cite{Babichev:2007dw}).
As can be easily checked, at leading order (over Minkowski space and in standard Cartesian coordinates) $V_{\pm}$ do indeed reduce to $c_s$. 
This can be done by replacing Eqs.~\eqref{gamma1}--\eqref{gamma3} into Eq.~\eqref{Vpm} [c.f. also Eq.~\eqref{deteffmet} below].

Even more importantly, ${\rm det}(\gamma^{\mu\nu})$ may in principle cross zero at some finite time and space location
during an evolution, as a result of which $V_{\pm}$ would no more be real and distinct, and strong hyperbolicity would be lost.
In more detail, if ${\rm det}(\gamma^{\mu\nu})$ becomes zero (positive), the system becomes
parabolic (elliptic). An example of such \textit{mixed type}~\cite{Stewart:2002vd} systems
is given by the Tricomi equation~\cite{Ripley:2019hxt,Bernard:2019fjb,Figueras:2020dzx}
\begin{eqnarray}
\partial^2_{t}\phi(t,r) + t \, \partial^2_{r}\phi(t,r) = 0~,
\label{trico}
\end{eqnarray}
which is hyperbolic for $t<0$ (with characteristic speeds $\pm \, (-t)^{1/2}$) and elliptic for $t>0$.

To prevent this breakdown of hyperbolicity, it is useful to
look at the eigenvalues of $\gamma^{\mu\nu}$
\be
\l_\pm = \frac12 \left[ \g^{tt} + \g^{rr} \pm \sqrt{\left(\g^{tt}-\g^{rr}\right)^2 + (2{\g^{tr}})^2}\, \right] \,, \label{lpm}
\ee
and even more importantly at its determinant
\begin{equation}
\det(\gamma^{\mu\nu}) = -\frac{1}{\alpha^2 g_{rr}} \left(1 +  \frac{2\, K''}{K'} X\right) \,. \label{deteffmet} \\
\end{equation}
From this expression it is clear that a Tricomi type evolution system can 
be avoided altogether if the function $K(X)$ defining the theory
 satisfies 
\be\label{condK} 
1 +  \frac{2\, K''(X)}{K'(X)} X >0 \,,
\ee
for all values of $X$ (see also \cite{Babichev:2007dw, Brax:2014gra}). If we take for instance $\beta=0$, this condition is indeed satisfied for any $\gamma>0$.
In the following we will focus therefore on this class of theories for simplicity, but
more generic values of $\beta$ and $\gamma$ can also enforce Eq.~\eqref{condK} for any value of $X$.
For  instance, in the units $\Lambda=1$ adapted to the problem
that we will employ later (c.f. Sec. \ref{units}),
for  $\gamma\sim 1$,
any (positive) $\beta\lesssim 1.2$ ensures that the condition \eqref{condK} is satisfied for all $X\in(-\infty,+\infty)$.

Having recognised that a Tricomi type behaviour can never occur for this class of K-essence theories,
let  us note, however, that  the characteristic speeds $V_\pm$ can in principle still
diverge. An example of this behaviour is provided by the Keldysh equation~\cite{Ripley:2019hxt,Bernard:2019fjb,Figueras:2020dzx}
\begin{equation}
t\partial^2_t\phi(t,r) + \partial^2_r\phi(t,r) = 0~,
\label{keld}
\end{equation}
which is hyperbolic (with characteristic speeds $\pm \, (-t)^{-1/2}$)
for $t<0$. In this case the system is strongly hyperbolic before the divergence of the characteristic speeds,
but  as the latter is approached the Courant-Friedrichs-Lewy stability condition 
requires any numerical integration to take smaller and smaller  steps (actually, infinitesimal steps when the characteristic speeds
diverge)~\cite{Bernard:2019fjb}. 
As we will show  in the following, we find that even for theories
satisfying Eq.~\eqref{condK}, a Keldysh type breakdown of
well-posedness can still occur outside black hole horizons, but
only  for initial data close  to critical black hole collapse.
We also stress  that superluminal or even diverging 
eigenvalues for the principal part of the evolution equations
are not necessarily
pathological from a physical viewpoint: for instance, the same can occur also in GR in certain gauges, e.g.
for a wave equation on flat space in Eddington-Finkelstein coordinates, ${\rm d}s^2=-{\rm d} v^2+2 {\rm d}v {\rm d}r+r^2 {\rm d}\Omega^2$.

\subsection{Regime of well-posedness}

Several papers in the literature~\cite{2011JHEP...04..096A,2011JHEP...10..028L,Bernard:2019fjb,Figueras:2020dzx} have shown that, in different situations, the evolution of an arbitrary initial data in K-essence features 
either a Tricomi or Keldysh type breakdown of the Cauchy problem.
However, a clear characterisation of when and why this happens is still missing. The main goal of this paper is to overcome some of these 
pathologies, and provide a recipe to identify critical regions where well-posedness (still) seems to break down.

First of all, let us put forward a few considerations about the linear/non-linear regime of K-essence.
Looking at the determinant (\ref{deteffmet}) and at the form of $K(X)$ given in (\ref{Kform}), it is not surprising that, as long as $X/\L^4 \ll 1$ during the whole evolution, the system of equations is globally well-posed. In the perturbative regime, higher powers of $X$ are for all purposes irrelevant.
Therefore, since they do not produce any sizeable effects (see discussion in Sec. \ref{screening}), they cannot  jeopardise well-posedness either. 
Conversely, in the non-linear regime $X/\L^4 \sim 1$, where these operators start producing interesting effects, the determinant (\ref{deteffmet}) may become zero and change the character of the scalar field evolution equations.

Since we are mostly interested in exploring the non-linear regime of $K(X)$ theories, the first requirement that one  needs to satisfy is
given by Eq.~\eqref{condK}, which prevents Eq.~\eqref{deteffmet} from vanishing altogether.
As already mentioned, this condition  (which can be enforced e.g. in theories with $\beta = 0$ and $\gamma=1$) 
protects the system from a Tricomi type change of character 
but not from a Keldysh type behaviour, which can (and indeed does) still arise.

Since  a Keldysh type  breakdown of well-posedness seems unavoidable, at least in our gauge\footnote{As already mentioned,
diverging characteristic speeds can in principle arise as a result of gauge choices. As we will comment in the following,
we have  unsuccessfully tried to prevent $V_\pm$ from diverging by introducing a non-zero shift.} and for analytic and Lorentz invariant $K(X)$ EFTs,
the most one can do is to  attempt to identify a ``regime of well-posedness'' for the initial data,
i.e. to characterise the class of initial data for which
the characteristic speeds $V_\pm$ remain real, finite and distinct.
 
To this purpose, we introduce a criterion inspired by the {\it hoop conjecture}, proposed by Kip Thorne in 1972~\cite{Klauder:1972je} for GR. This conjecture states that a configuration of energy $E$  forms a black hole
horizon when and only when it occupies a region whose circumference $C$ in every
direction is such that $C/(G  E) \leq 4\pi$.

Similarly, we characterise the initial configuration of the scalar field by 
its ``compactness'' $\CC= G\E/\s$, defined as the ratio of the configuration's energy $\E$ and its ``width''~$\s$.
We will define $\E$ and $\s$ more precisely in the following, but for the moment
let us mention that our simulations are consistent with
the hypothesis that it is the compactness of the initial data that
defines whether the evolution is well-posed. In more detail, in Sec. \ref{results}
we find that 
if $\CC \gtrsim 0.39$ the evolution  leads to a black hole and is well-posed outside apparent black hole horizons,
while for $\CC \lesssim 0.37$ the evolution is again well-posed, but the scalar field disperses away and the geometry relaxes to flat space.
Note that these conclusions are independent of whether the scalar field is in the linear or non-linear regime:
 we will show cases where $X/\L^4 \lesssim 1$ and  a black hole still forms without any breakdown of the Cauchy problem outside the horizon.
Interestingly, the range of values for which a Keldysh type change of character occurs (and thus well-posedness is lost)
is given by the very narrow interval $ 0.37 \lesssim \CC \lesssim 0.39$ at the transition between Minkowski and black hole formation (i.e.~near critical black hole collapse).

%%%%%%%%%%%%%%%%%%%%%%%%%%%%%%%%%%%%%%%%%%%%%%%%%%%%%%%%
\section{Numerical setup}
\label{NumSetup}
%%%%%%%%%%%%%%%%%%%%%%%%%%%%%%%%%%%%%%%%%%%%%%%%%%%%%%%%

\subsection{Units}
\label{units}

Unlike GR or Fierz-Jordan-Brans-Dicke like theories, our K-essence action (\ref{action})-(\ref{Kform}) is characterised by the strong-coupling scale $\L$,
which in our units $c=1$ has dimensions $[E/L^3]^{1/4}$, with $E$ an energy and $L$ a length. To perform
numerical simulations, it is convenient to choose units suited for the problem, namely ones
in which $G=c=\L=1$. This corresponds to measuring times in units of 
$c\,G^{-1/2} \L^{-2}$, lengths in units
of $c^2\,G^{-1/2} \L^{-2}$, masses in units of $c^4\,G^{-3/2} \Lambda^{-2}$, etc.
By using these geometrised units, the results of our simulations can be rescaled to any K-essence theory with
 action (\ref{action})-(\ref{Kform}), irrespectively of the value of $\L$. Clearly, when
rescaling to different theories, the mass/energy $\E$ and width $\s$ of the initial data will be different,
but the compactness $\CC=G\E/\s$  remains invariant. As a result, our conclusions for the range 
of $\CC$ yielding well-posed evolutions apply to generic $K(X)$ theories of the  form (\ref{action})-(\ref{Kform}).

\subsection{Implementation details and code}

The numerical code used in this work is an extension of the one presented in 
Ref.~\cite{Alic:2007ev} for one dimensional black hole simulations, and used in Ref.~\cite{Bernal:2009zy,Marsh:2015wka,Raposo:2018rjn} to study the dynamics of boson stars, neutron stars, fermion-boson stars and anisotropic stars. 

It uses a high-resolution shock-capturing finite-difference (HRSC) scheme, described in Ref.~\cite{Alic:2007ev}, to discretise the spacetime variables. In particular, this method  can be viewed as a fourth-order finite difference scheme plus third-order adaptive dissipation. The dissipation coefficient is given by the maximum propagation speed at each grid point. 
For the scalar field, we use a more robust HRSC second-order method, namely the Local-Lax-Friedrichs flux formula  with a monotonic-centred limiter~\cite{CCC,Palenzuela:2018sly}.
The time evolution is performed through the   method of lines using a third-order accurate strong stability preserving Runge-Kutta integration scheme, with a Courant factor of $\Delta t/\Delta r = 0.25$, so that the Courant-Friedrichs-Levy condition imposed by the principal part of the system of equations is satisfied. Most of the simulations presented in this work have been performed with a spatial resolution of $\Delta r = 0.01$, in a domain with outer boundary located at $r= 120$ (in the units of Sec.~\ref{units}). We have verified that the results do not vary significantly when the position of the outer boundary is changed. We use maximally dissipative boundary conditions for the spacetime variables, and outgoing boundary conditions for the scalar field.

\subsection{Diagnostic tools}

In this section, we provide a more precise definition of the  ``compactness'' $\CC$ of the initial data, and we 
present diagnostic tools to identify the presence of apparent black hole and sound horizons.

To define the compactness of the initial data, we fist need to characterise the energy of the scalar field configuration. A simple
estimate can be obtained as the spatial  integral of the  energy density of the scalar field,  $\tau_{\phi} = n_\mu n_\nu T_{\phi}^{\mu\nu}$, i.e.
\begin{eqnarray}
\E &=& \int dr 4 \pi r^2 \sqrt{g_{rr}} g_{\theta\theta}  \tau_{\phi}  \nonumber~, \\
&=& 4 \pi \int dr \left[ r^2 \sqrt{g_{rr}} g_{\theta\theta} (- K  - 2 K' \Pi^2)\right] \,.\label{ener}
\end{eqnarray}
As for the size, since our initial data for the scalar field will consist of Gaussian pulses, we 
utilise their root mean square (r.m.s.) width as a measure of $\s$. We have checked that the width of the pulses changes during the simulations by at most a factor $2$,
which is sufficient for our order-of-magnitude estimates. The compactness of our initial scalar field configuration is then defined as 
\be
\CC \equiv \frac{\E}{\s} \,.
\label{comp}
\ee

Throughout the evolution, the formation of a black hole is signalled  by the appearance of an apparent horizon~\cite{Thornburg:2006zb}. 
By definition, the latter is the
outermost trapped surface, i.e.~the largest two-surface where the outgoing null ray congruence has vanishing expansion $\Theta$.
In terms of 3+1 quantities, the apparent horizon is then defined by the condition 
\begin{equation}
\Theta=\nabla_{k}s^{k} + \K_{ij}s^{i}s^{j}- \mathrm{tr}\K=0\,,
\label{expAH1}
\end{equation}
where $s^{i}$ is the outgoing unit normal to the apparent horizon. In spherical symmetry, the expansion reduces to
the algebraic expression
\begin{equation}
\Theta=\frac{1}{\sqrt{g_{rr}}}\left( 2{D_{rr}}^{r} +\frac{2}{r} \right) - 2{\K_{\theta}}^{\theta}~.
\label{expAH2}
\end{equation}
Therefore, in order to find the apparent horizon, we look for the (outermost) change of sign in the expansion~\eqref{expAH2}, 
which corresponds physically to a ``collapse'' of the outgoing null ray congruence. One can show that an
equivalent criterion~\cite{Nielsen:2005af,Faraoni:2016xgy} (c.f. also Eq. 6.2 of \cite{Baumgarte:2002jm}) consists of looking for zeros of the contravariant metric component $g^{\tilde{r} \tilde{r}}$, where  $\tilde{r}(r,t)$ is the areal\footnote{The name ``areal radius'' is due to the fact that in a given hypersurface, a fixed coordinate  $\tilde{r}$ defines a two dimensional sphere with  surface area $4\pi\tilde{r}^2$.} (or polar) radius, i.e. the radial coordinate in terms of which the angular part of the line element becomes $\tilde{r}^2{\rm d}\Omega^2$.

In addition, in K-essence 
one can define also an effective horizon for the scalar field perturbations. This is referred to as ``sound horizon'' in
the literature~\cite{Babichev:2007dw}. For a static effective metric $\gamma^{\mu\nu}$, the sound horizon
corresponds to a zero of $\gamma_{tt}$ (or equivalently of $\gamma^{rr}$). For time dependent metrics, 
one can define an apparent sound horizon in terms of zeros of the expansion of the outgoing 
null rays of $\gamma_{\mu\nu}$, or equivalently in terms of zeros of $\gamma^{\tilde{r}\tilde{r}}$, with $\tilde{r}(t,r)$ again the areal radius. The corresponding
condition can be shown to be
\begin{eqnarray}
\mathcal{S} &=& r^2 g_{\theta\theta} \left[(r {D_{r\theta}}^{\theta}+2)^2 \gamma^{rr}\right. \label{expSH} \\  
&+& \left.  r {\K_{\theta}}^{\theta} \alpha (r \gamma^{tt} {\K_{\theta}}^{\theta} \alpha -2 (r {D_{r\theta}}^{\theta}+2)\gamma^{tr})\right] = 0\,. \nb
\end{eqnarray}

\subsection{Initial data}

Equilibrium configurations for non-rotating, self-gravitating objects in spherical symmetry can be obtained by imposing the condition for stationary  solutions, $\partial_{t}{\bf U}= 0$.  The resulting equations can be further simplified by using the maximal polar slicing condition $\K^{\theta}\,_{\theta}=0$. The line element
employs polar (areal) coordinates and reads
\begin{equation}
ds^{2}=-\alpha^{2}(\tilde{r})dt^{2}+a^{2}(\tilde{r})d\tilde{r}^{2}+\tilde{r}^{2}d\Omega^{2}~,
\label{polar_metric}
\end{equation}
where $d\Omega^{2}= d\theta^{2} + \sin^{2}\theta d\varphi^{2}$ is the metric of a unit two-sphere.
In these coordinates, the $tt$ and $\tilde{r}\tilde{r}$ components of Eq.~\eqref{EE} reduce to a set of ordinary differential equations, which can be solved for a given scalar field profile.
For the latter, we consider
\begin{eqnarray}
\phi(t=0,\tilde{r})&=& A\exp\left[- \frac12 \left(\frac{\tilde{r}-\tilde{r}_0}{\s}\right)^2\right] \text{Sin}\left[\frac{\tilde{r}-\tilde{r}_0}{\sqrt{2}\,  \s}\right] \,, \label{ICphi} \nonumber \\[2ex]
\Phi(t=0,\tilde{r})&=& \partial_{\tilde{r}}\phi \,, \qquad \Pi(t=0,\tilde{r})= 0 \,,
\end{eqnarray}
with $\tilde{r}_0 = 55$, $A$ varying between $0.01$ and $0.05$, and $\s$ varying between $2$ and $6$.
These initial data correspond to scalar field compactnesses $\CC$ spanning the range between $0.03$ and $2.37$.

In order to obtain  physically meaningful solutions, appropriate boundary conditions have to be  imposed
to ensure regularity at the origin and asymptotic flatness at large distances, namely
\begin{eqnarray}
a(\tilde{r}=0) &=& 1 \,, \qquad \alpha(\tilde{r}=0) = 1\,, \nb \\ a'(\tilde{r}=0) &=& 0 \,, \qquad \alpha'(\tilde{r}=0) = 0 \,, \label{bc1}  \\ 
\lim_{\tilde{r}\to\infty}a(\tilde{r}) &=& 1 \,,  \qquad \lim_{\tilde{r}\to\infty}\alpha(\tilde{r}) = {\rm const}\,. \nb  
\label{bc2}
\end{eqnarray} 
Taking these boundary conditions into account, the initial data can be solved for by integrating from $\tilde{r}=0$ toward the exterior boundary $\tilde{r}=\tilde{r}_{out}$ using a 
 fourth order Runge-Kutta integrator. After the integration  of the equilibrium configuration,
it is always possible to rescale~$t$ so that $\lim_{\tilde{r}\to\infty}\alpha(\tilde{r}) = 1$, at the price 
of having  $\alpha(\tilde{r})$ different from zero at the centre.

Once the solution for the initial data has been found, a coordinate transformation is performed from areal coordinates to maximal isotropic ones, in which the line element is
\begin{eqnarray}
ds^{2} = -\alpha^{2}(r) dt^{2} + \psi^{4}(r)(dr^{2} + r^{2}d\Omega^{2})~,
\label{metric_conformal}
\end{eqnarray}
where $\psi$ is the conformal factor. This line element is then mapped into the metric ansatz used for the evolution, i.e. Eq.~\eqref{metric_ansatz}.

\section{Results}
\label{results}
In this section, we summarise the outcome of the numerical simulations that we performed and which allowed us to identify the regime of well-posedness of K-essence theories.
We use the parameters $\beta = 0$ and $\gamma=1$ in the $K(X)$ action~(\ref{Kform}). This choice, besides preventing the determinant (\ref{deteffmet}) from ever becoming zero [c.f. Eq.~\eqref{condK}], also allows for the kinetic screening discussed in Sec. \ref{screening}.
We have performed evolutions with different resolutions, which indicate that the results presented here are robust and show second order convergence.

\begin{table}
	\begin{ruledtabular}
		\begin{tabular}{cc||cc|c|c}
			$A$ & $\sqrt{2}\, \sigma$ & $\E$ & $\CC$ & Final fate  &  $\max |X/2|$  
			\\ \hline\hline 
			0.001   & 2   & 0.008  & 0.058   & Minkowski &   0.00045 \\
			0.005   & 2   & 0.207  & 0.146   & Minkowski &   0.0175 \\
			0.007   & 2   & 0.4    & 0.280   & Minkowski &   0.09   \\
			0.0072  & 2   & 0.428  & 0.303   & Minkowski &   0.1125 \\
			0.0076  & 2   & 0.477  & 0.337   & Minkowski &   0.18   \\			
			0.008   & 2   & 0.52   & 0.37    & Diverge vel. &  0.57     \\
			0.0082  & 2   & 0.555  & 0.39    & Black Hole$^{\star}$&  1.4    \\
			0.0086  & 2   & 0.61   & 0.43    & Black Hole$^{\star}$&  0.53   \\
			0.009   & 2   & 0.668  & 0.47    & Black Hole          &  0.7    \\
			0.011   & 2   & 0.908  & 0.64    & Black Hole          &  0.6    \\
			0.012   & 2   & 1.18   & 0.83    & Black Hole          &  0.225  \\
			0.013   & 2   & 1.3    & 0.98    & Black Hole          &  0.17      \\
			0.014   & 2   & 1.604  & 1.14    & Black Hole  &  0.1    \\
			0.015   & 2   & 1.8    & 1.2     & Black Hole  &  0.2       \\
			0.016   & 2   & 2.08   & 1.47    & Black Hole  &  0.2    \\ 
			0.018   & 2   & 2.6    & 1.85    & Black Hole  &  0.045  \\ 
			0.0185  & 2   & 2.7    & 1.96    & Black Hole  &  0.027  \\
			0.019   & 2   & 2.9    & 2.06    & Black Hole  &  0.022  \\
			0.0195  & 2   & 3.06    & 2.17   & Black Hole  &  0.021  \\ 			 
			0.02    & 2   & 3.22   & 2.37    & Black Hole  & 0.01 \\
			\hline\hline
			0.05   & 4   & 0.103   & 0.036    & Minkowski  &  0.007 \\
			0.014  & 4   & 0.8     & 0.286    & Minkowski  &  0.023 \\
			0.015  & 4   & 0.99    & 0.35     & Minkowski  &  0.06  \\
			0.016  & 4   & 1.05    & 0.37     & Minkowski  &  0.2 \\	
			0.0161 & 4   & 1.068    & 0.377   & Minkowski  &  0.48  \\					
			0.0163 & 4   & 1.09    & 0.387    & Diverge vel. &  0.115 \\
			0.0165 & 4   & 1.12    & 0.39     & Black Hole$^{\star}$ &  0.89 \\
			0.0175  & 4  & 1.26    & 0.44     & Black Hole &  0.4\\
			0.018  & 4   & 1.33    & 0.47     & Black Hole &  0.28\\
			0.0185 & 4   & 1.4     & 0.49     & Black Hole &  0.37 \\
			0.019  & 4   & 1.48    & 0.52     & Black Hole &  0.5 \\									
			0.0195 & 4   & 1.56    & 0.55     & Black Hole &  0.75 \\									
			0.02   & 4   & 1.64    & 0.58     & Black Hole &  1.49 \\
			\hline\hline
			0.01  & 6  & 0.27   & 0.06   & Minkowski   &  0.006 \\
			0.02  & 6  & 1.16   & 0.259  & Minkowski   &  0.007 \\
			0.021 & 6  & 1.21   & 0.28   & Minkowski   &  0.01 \\
			0.023 & 6  & 1.45   & 0.34   & Minkowski   &  0.02 \\
			0.024 & 6  & 1.58   & 0.37   & Minkowski   &  0.12\\
			0.0241 & 6  & 1.59   & 0.375   & Minkowski   &  0.21 \\
			0.0243 & 6  & 1.61   & 0.381   & Black Hole   & 0.9 \\
			0.025 & 6  & 1.71   & 0.40   & Black Hole   &  0.3 \\
			0.026 & 6  & 1.85   & 0.43   & Black Hole   &  1.34 \\
			0.028 & 6  & 2.14   & 0.5    & Black Hole   &  0.2 \\
			0.03  & 6  & 2.45   & 0.577  & Black Hole  &  0.6  \\
			\hline\hline
		\end{tabular}
		\caption{{\em Numerical simulations.} The entries of the table are respectively: the amplitude and r.m.s. width of the pulse~\eqref{ICphi}, the energy~\eqref{ener} and compactness~\eqref{comp} of the scalar field initial profile, the outcome of the simulation and the maximum of~$|X/2|$. Note that the cases reported here are merely a subset of all the simulations that we ran.
		}\label{casestab}
	\end{ruledtabular}
\end{table}

\begin{figure}
	\centering
	\includegraphics[width=0.5\textwidth]{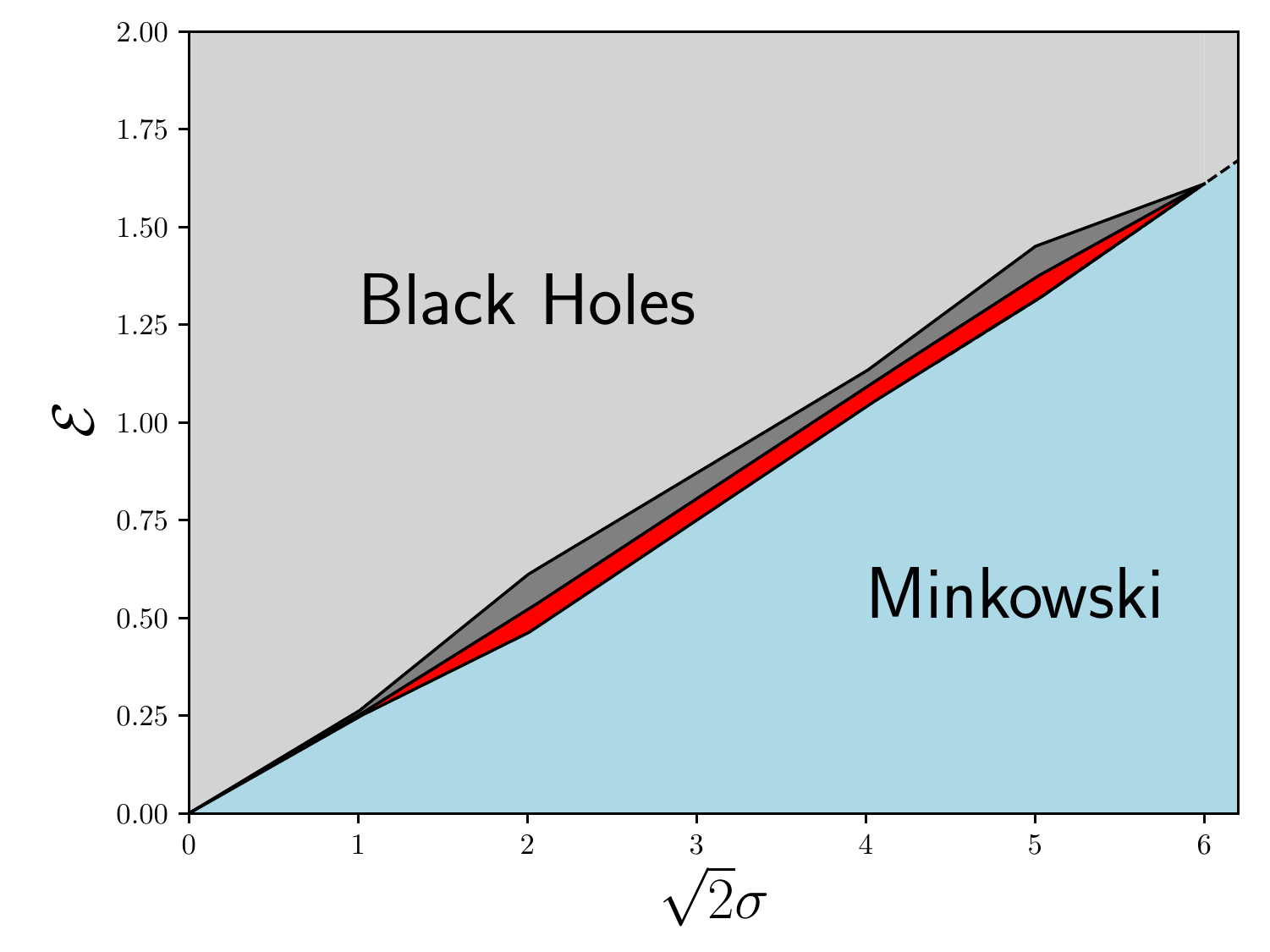}
	\caption{{\em Dynamics of the scalar field and final state,} as a function of energy $\E$ and width $\s$ of the initial  scalar field profile.
The light blue region ($\CC \lesssim 0.37$) denotes the parameter space where the scalar field disperses away and the evolution
asymptotes to Minkowski space, while in the light-grey region ($ \CC \gtrsim 0.43$) 
a black hole forms. In the dark-grey region ($0.39 \lesssim \CC \lesssim 0.43$), an apparent sound horizon
forms inside an apparent black hole horizon; this is a situation which cannot be followed in our gauge (which cannot
capture the formation of black hole and sound horizons at the same time), but which is not pathological.
Finally, in the red region ($0.37 \lesssim \CC \lesssim 0.39$) the evolution
produces diverging characteristic velocities for the scalar field, i.e. a Keldysh type change of character
of the evolution equations.}
\label{fig:validity} 
\end{figure}

Our main results are summarised in Table \ref{casestab} and shown in Fig.~\ref{fig:validity}. We report the different final states obtained from the dynamical evolution of the system, as a function of the compactness $\CC$ of the initial data. In all cases, the scalar field initially moves toward the origin, increasing its amplitude approximately as $1/r$, and the evolution leads to one of the following four possible scenarios: 
\begin{itemize}
	\item For low values of the compactness $\CC \lesssim 0.37$,  the scalar field grows as it reaches the origin and then disperses to infinity, leaving a flat Minkowski spacetime. The evolution is always stable and well-posed.
	\item For values of the compactness near but below the black hole formation threshold, $0.37 \lesssim \CC \lesssim 0.39$, 
the characteristic speeds of the scalar field diverge as the latter moves toward the centre, giving rise to
a Keldysh type change of character of the evolution equations. As stressed previously, 
diverging eigenvalues for the principal part of the evolution equations are not pathological per se, e.g. they occur also in GR due to gauge effects. 
We have indeed verified that by introducing a non-vanishing shift in our metric ansatz \eqref{metric_ansatz},
those eigenvalues do indeed become finite, but we could not identify an
evolution equation for the shift yielding a strongly hyperbolic system, due to the highly non-linear nature of the problem.
Trial and error attempts for the shift driver seemed to produce evolutions that eventually
become unstable on small spatial scales.
Irrespective of its conceptual meaning (physical or related to the gauge choice), the 
divergence of the principal part's eigenvalues is problematic in practice, as it causes 
simulations to grind to a halt as a result of the 
Courant-Friedrichs-Lewy stability condition. A possible solution would be to utilise an implicit time integrator,
at the expenses of losing small-timescale details of the solution (and potentially their cumulated secular effects, if any).
	\item 
For values of the compactness near but above the black hole formation threshold, $0.39 \lesssim \CC \lesssim 0.43$, the scalar field collapses and forms a black hole. First an apparent horizon is formed, and soon after an apparent sound horizon. The inability of our gauge to capture both horizons simultaneously leads to a failure of our simulations after  black hole formation. However, this scenario could be prevented by excising the interior of the horizon from our computational domain~\cite{Ripley:2019aqj}, or possibly by using more involved gauge conditions on the lapse and shift. 	Our attempts to identify
such gauge conditions, however, have been unfruitful. 
	\item For larger values of the compactness, $ \CC \gtrsim 0.43$, the system collapses to a black hole without forming immediately  a sound horizon, which is well hidden in the interior of the apparent horizon. We can simulate the black hole formation and the subsequent evolution without encountering any breakdown of the Cauchy problem.
\end{itemize}

Note that the second and third scenarios arise in tiny ranges for the compactness. Finding these ranges in practice requires an accurate fine tuning between the amplitude and width of our initial pulse. At large widths $\sqrt{2}\s\sim 6$,
we could not identify initial data giving evolutions falling in the second and third scenarios, although they might  well exist.

Let us now comment on the linear/non-linear regime of the scalar field during the evolution. In the last column of Table \ref{casestab}, we report the maximum (absolute) value of the kinetic term $|X/2|$ during the evolution. Note that at the initial time the maximum of $|X/2|$ is always smaller than~$10^{-5}$, as larger values of $|X/2|$ (corresponding to even more compact initial data) would
lead to prompt black hole collapse. 
Inspection of this column shows
that most of our simulations are in the mildly non-linear regime.
Only for very low values of $\CC$ the scalar field evolution remains always linear at all times and the metric close to Minkowski. The highest non-linearities never exceed $|X/2| \sim 1.5$. This seems to suggest that (at least in vacuum) when the dynamical evolution of K-essence is well-posed, strong non linearities ($X/\L^4 \gg 1$) never appear in the dynamics, in agreement with standard EFT arguments.

Finally, let us mention that
the different behaviours outlined above were already observed in the literature~\cite{Bernard:2019fjb,Figueras:2020dzx}.
Ref.~\cite{Bernard:2019fjb}, however, could not follow the evolution of the black hole after the collapse, due to their choice of areal coordinates (which become singular when a horizon forms). Ref.~\cite{Figueras:2020dzx} 
attributed the problems that we described above and which take place near the black hole formation threshold
to the onset of critical collapse, and they found them for all values of the coupling constants in $K(X)$ (i.e. even for $\beta=\gamma=0$,
corresponding to a minimally coupled scalar field). Unlike Ref.~\cite{Figueras:2020dzx}, we
observed no pathologies in the limit $\beta=\gamma=0$.
 
\begin{figure*}
	\includegraphics[width=1.0\textwidth]{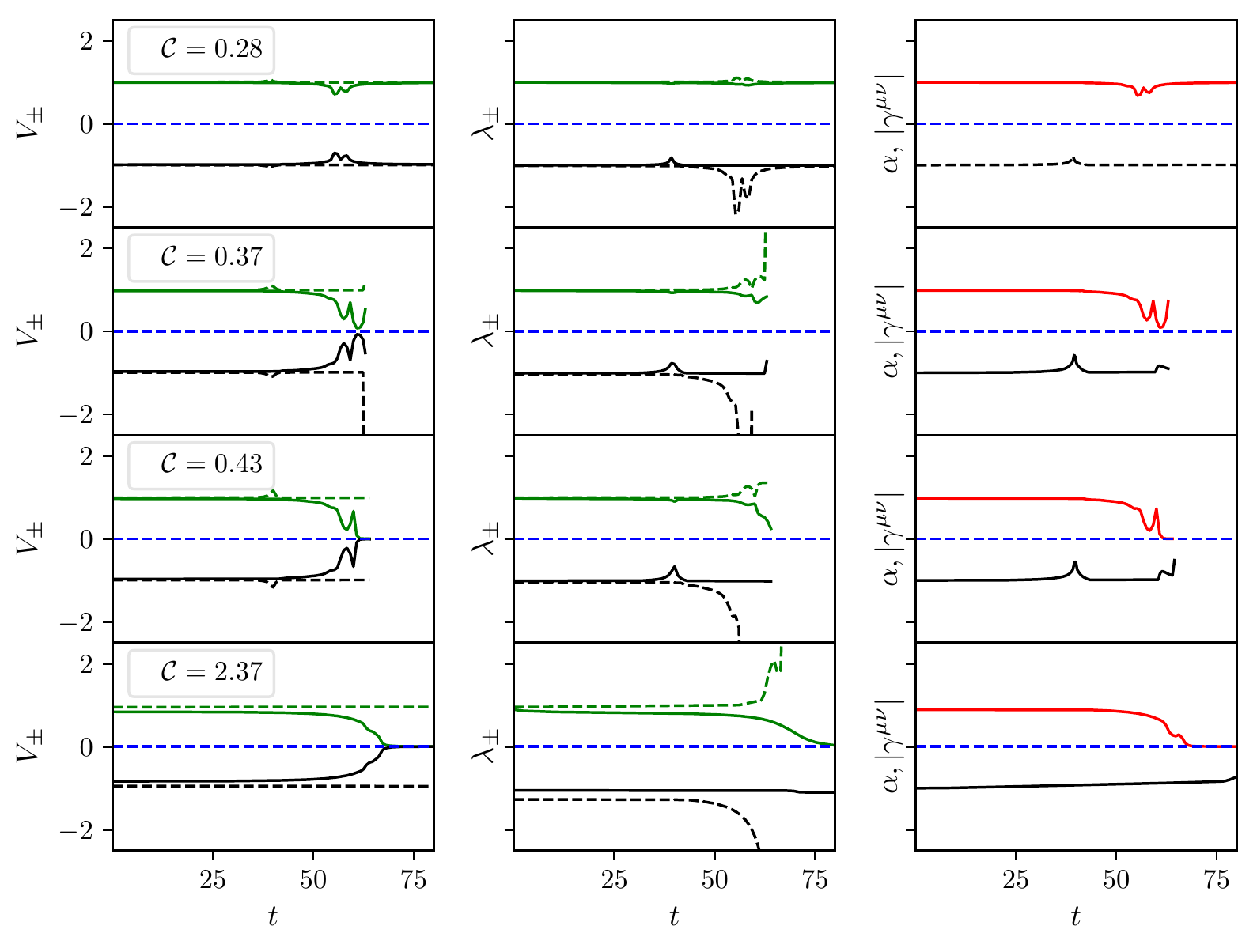}
	\caption{{\em Dynamics of the pulse.} Each row corresponds to a different compactness $\CC$ of the initial scalar field pulse (from top to bottom, $0.28, 0.37, 0.43$ and $2.37$). 
{\em First and second columns}: the green dashed line represents
	$\max(V_{+})$ and $\max(\lambda_{+})$; the green solid line is $\min(V_{+})$ and $\min(\lambda_{+})$; the black solid line is $\max(V_{-})$ and $\max(\lambda_{-})$; and the black dashed line is $\min(V_{-})$ and $\min(\lambda_{-})$. {\em Third column:} the solid red line is $\min{(\alpha)}$ (which is useful to track black hole formation), while the solid black line is $\max{\det(\gamma^{\mu\nu})}$ [which is always non-zero, c.f. Eq.~\eqref{condK}].
	}
	\label{fig:evolution}
\end{figure*}

\subsection{Examples of dynamical evolutions}
We now describe in more detail representative cases for each of the four scenarios
discussed above. Results for these evolutions  are shown in Fig.~\ref{fig:evolution}. 
In order to track the character of the scalar field equation, the maximum and minimum 
(on the spatial grid) of 
the eigenvalues of the principal part [Eq.~\eqref{Vpm}] and of the effective metric [Eq.~\eqref{lpm}] are displayed as a function of time.
Furthermore, the maximum of the determinant of the effective metric~\eqref{deteffmet} is
also plotted,
in order to show that it never becomes zero. [Since $\det(\gamma^{\mu\nu})<0$, a vanishing value for $\max{\det(\gamma^{\mu\nu})}$ would signal a Tricomi type
change of character, and which cannot happen by construction thanks to the condition \eqref{condK}].
Finally, to track the formation of a black hole, we plot the minimum of the lapse on the spatial grid as a function of time.

\begin{figure}
	\centering
	\includegraphics[width=0.5\textwidth]{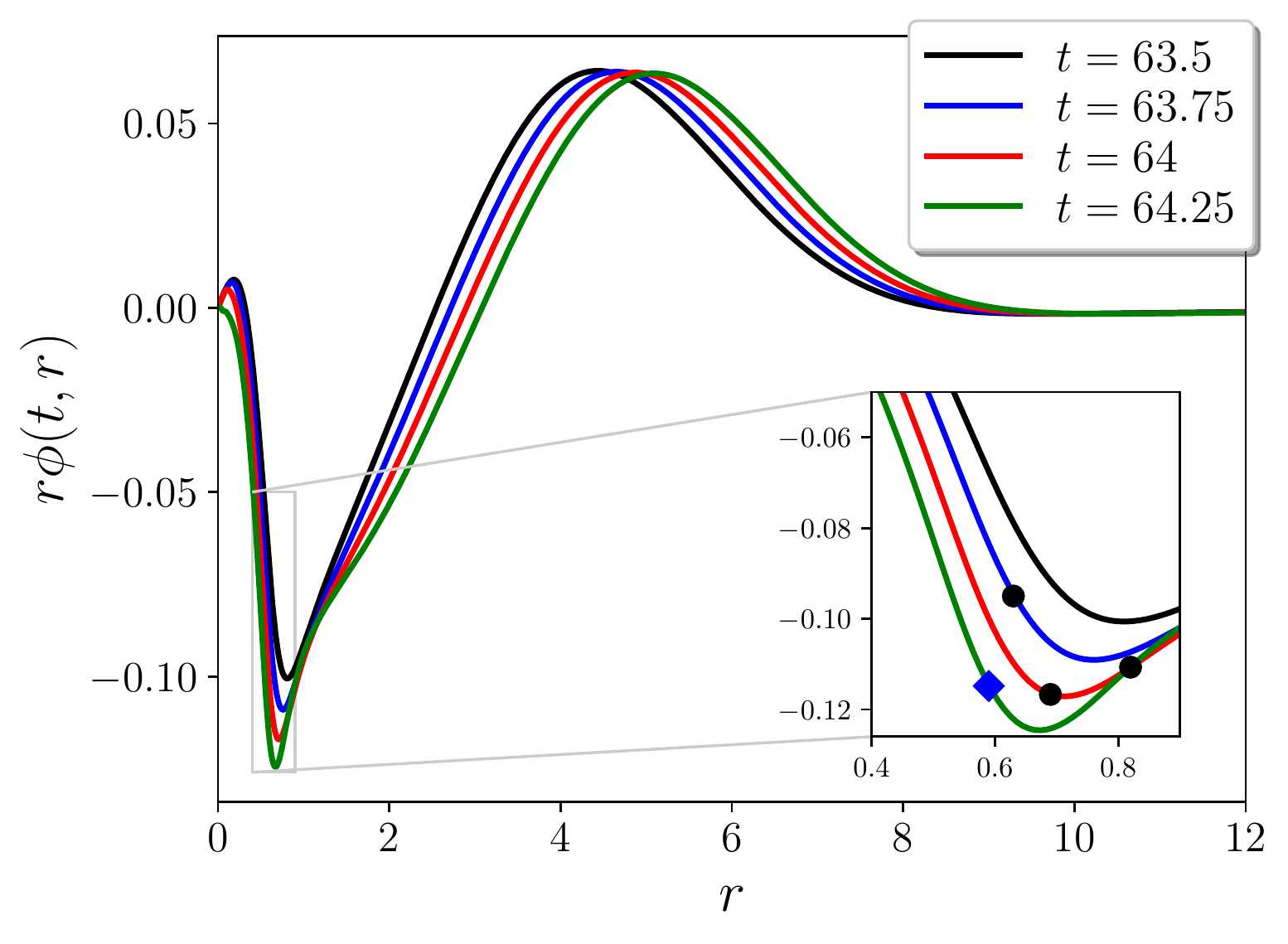}
	\caption{{\em Evolution of the scalar field for $\CC=0.43.$} Time snapshots of  the radial profile of the scalar field $\phi(t,r)$ (multiplied by $r$) near the collapse to a black hole. The black dots  denote the position of the apparent black hole horizon, while the blue diamond point marks the position of the (apparent) sound horizon, which appears shortly after.    
	}
	\label{fig:evolA0.0086}
\end{figure}

The four examples that we show are the following:
\begin{enumerate}
	\item[(i)] \bm{$\CC = 0.28.$} In this case the characteristic velocities [Eq.~\eqref{Vpm}] 
and the eigenvalues of the effective metric [Eq.~\eqref{lpm}] remain almost constant with opposite signs during the evolution, showing that the character of the scalar field equation is strongly hyperbolic. This behaviour is not surprising, since the gradients of the scalar field remain always very small during the evolution. The dynamics of the scalar field therefore effectively reduces to a wave equation in flat spacetime. Finally, the lapse remains constant and close to one, confirming that the metric is almost flat during the evolution.
	
	\item[(ii)]\bm{$\CC = 0.37.$} This case shows an exponential growth of the characteristic speeds [Eq.~\eqref{Vpm}]. This indicates that the scalar field equation behaves like a Keldysh one. Note that there is no formation of an apparent or sound horizon, since the expansions 
for the null rays of $g_{\mu\nu}$ and $\gamma_{\mu\nu}$ [c.f. Eqs.~\eqref{expAH2} and \eqref{expSH}]
remain always positive, even when the lapse is close to zero.

	\item[(iii)] \bm{$\CC = 0.43.$} In this case the system collapses to a black hole. This can be seen from the minimum of the lapse becoming zero. In Fig.~\ref{fig:evolA0.0086} we display the evolution of the scalar field near the collapse to a black hole. (Note that because of our gauge choice, the pulse freezes inside the black hole). The black dots indicate the position of the apparent black hole horizon (which first appears at $t\approx 63.75$) and the blue diamond  marks the position of the (apparent) sound horizon, which forms right after the black hole (apparent) horizon.
As mentioned already, 
our gauge cannot capture both horizons simultaneously, and the simulation thus crashes 
after the sound horizon formation. 
As observed in~\cite{Ripley:2019aqj,Figueras:2020dzx},
since this happens inside the black hole, which is causally disconnected from the exterior, there is no breakdown of the Cauchy problem.
This behaviour may be due to the formation of large shocks (since the scalar field has a very steep profile
that produces large gradients), and it may be avoided by excising the interior of the horizon from our computational domain~\cite{Ripley:2019aqj},
or by devising  more involved gauge conditions  on the lapse and shift.

	\item[(iv)]\bm{$\CC = 2.37.$} This last case shows the collapse of the pulse into a black hole, as can be seen from the minimum of the lapse going to zero. Here, the characteristic speeds of the principal part [Eq.~\eqref{Vpm}] and the eigenvalues of the effective metric [Eq.~\eqref{lpm}] remain well-behaved before the black hole is formed. Note that as the black hole is formed, the eigenvalues can approach zero, but that is simply 
due to the lapse going to zero and to our gauge choice that freezes the pulse evolution.
\end{enumerate}

\section{Conclusions}
\label{Conc}

In this paper, we have studied the well-posedness of the Cauchy problem in 
K-essence theories, i.e. nowadays the most compelling alternative to Dark Energy.
Focusing on the spherically symmetric case, we have identified a large class of theories,
defined by the condition~\eqref{condK}, for which a Tricomi type change of character
of the evolution equations (from hyperbolic to parabolic) can never take place.
Furthermore, even though caustics/shocks are known to naturally appear in K-essence,
we have introduced a conservative formulation for the scalar field equation that allows for
using techniques from computational fluid dynamics to resolve the shocks.

Thanks to these improvements over previous work~\cite{Ripley2019,Ripley:2019aqj,Ripley:2019hxt,Bernard:2019fjb,Figueras:2020dzx}, we are  able to
solve the vacuum Cauchy problem in spherical symmetry
without any breakdown of well-posedness, at least outside (apparent) black hole horizons and
except for a small set of initial data very close to critical black hole collapse.
For these initial data, 
which we  describe in terms of the ``compactness'' of the 
initial scalar field profile, 
 the (coordinate) characteristic speeds of the scalar field diverge, leading to
a Keldysh type change of character of the evolution system. 

This divergence of
the coordinate characteristic speeds is not physically pathological in itself (e.g. it can appear also
in GR on a Minkowski background if one describes the latter in 
Eddington-Finkelstein coordinates), but presents at the very least a practical problem,
as the time-step of a Cauchy problem integrator becomes infinitesimally small
because of causality (i.e the  Courant-Friedrichs-Lewy stability condition).
In this sense, Keldysh type change of character that we find
makes the theory lose predictability (at least as far as the Cauchy problem is
concerned).
By trial and error, we
managed to find examples of gauges (with non-zero shift) where the
divergence of the  characteristic velocities disappears, but those gauges do not seem to provide a strongly hyperbolic system (which
leads again to a breakdown of the Cauchy problem and predictability). 
Another approach to 
deal with such a Keldysh-type breakdown of well-posedness would be to 
use an implicit integrator, at the price of neglecting the dynamics on short timescales (including
their secular effects, if there are any). We
will develop both these possibilities in future work.

Remarkably, during all of our evolutions 
the highest non-linearities never exceed  $X/\L^4 \sim 1.5$, suggesting that well-posed  evolutions in K-essence can never develop strong non-linearities, at least in vacuum.
This is in agreement with standard EFT arguments. It is not to be excluded, however, that suitable
non-linear and/or UV completions may fix the Keldysh type change of character of the evolution system that we find. It should be noticed, however,
that standard UV completions may not exist for the class of K-essence theories that 
gives a screening mechanism, because of positivity bounds~\cite{Adams:2006sv}.

Finally, we note that our results also set the stage to perform numerical simulations of the 1+1 and 3+1 dynamics  in the presence of realistic matter sources \cite{terHaar:2020xxb}.

% % % % % % % % % % % % % % % % % % % % % % % % % % % % 
% % % % % % % % % % % % % % % % % % % % % % % % % % % % 

\subsection*{Acknowledgments} 
We thank Luis Lehner and Lotte ter Haar for many enlightening discussions.
M.B., M.C. and E.B. acknowledge financial support provided under the European Union's H2020 ERC Consolidator Grant
``GRavity from Astrophysical to Microscopic Scales'' grant agreement no. GRAMS-815673.
C.P. acknowledges support from the Spanish Ministry of Economy and Competitiveness grants AYA2016-80289-P and PID2019-110301GB-I00 (AEI/FEDER, UE).

% % % % % % % % % % % % % % % % % % % % % % % % % % % % 
\bibliographystyle{utphys}
\bibliography{biblio}
 
\end{document}